\documentstyle[epsfig]{elsart}

\begin{document}
\begin{frontmatter}

%
\title{Analyses of multiplicity distributions at Tevatron by a two-component 
stochastic model\\ --- No leading particle effect in E735 Experiment ---}

\author[Shinshu]{M. Biyajima,}
\author[Toba]{T. Mizoguchi,}
\author[RCNP]{N. Nakajima,}
\author[ICRR]{A. Ohsawa}
\author[Matsu]{and N. Suzuki}

\address[Shinshu]{Department of Physics, Faculty of Science, Shinshu 
University, Matsumoto 390-8621, Japan}
\address[Toba]{Toba National College of Maritime Technology, Toba 517-8501, 
Japan}
\address[RCNP]{Research Center for Nuclear Physics, Osaka University, Ibaraki 
567-0047, Japan}
\address[ICRR]{Institute for Cosmic Ray Research, University of Tokyo, 
Kashiwa, 277-8582, Japan}
\address[Matsu]{Matsusho Gakuen Junior College, Matsumoto 390-1295, Japan}

%
\begin{abstract}
Comparisons of multiplicity distributions at $\sqrt s = 1.8$ TeV (E735 
Experiment and CDF Collaboration) with our predictions by a two-component 
stochastic model have been made. This stochastic model is described by the 
pure-birth and Poisson processes. It is found that there are discrepancies 
among data at the CERN S\= ppS collider and those at the Tevatron collider. 
The latter data by E735 do not contain the leading particle effect described 
by the Poisson process, providing that the view of the two-component 
stochastic model is correct. The data by CDF contain small leading particle 
effect. The reason is probably attributed to the correction made by E735, 
which is necessary to make the data of the full phase space from the 
restricted rapidity $|\eta|<3.25$.
\end{abstract}
\end{frontmatter}

%
\section{Introduction}
In 1993, we made predictions for multiplicity distributions at the Tevatron 
collider by a two-component stochastic model including the pure-birth (PB) 
process and the Poisson process \cite{mizo93}. See also \cite{biya91}. Those 
predictions have been based on analyses of $C_q = \langle n^q \rangle/\langle 
n \rangle^q$ ($q = 3\sim 5$) in the energy range $\sqrt s = 11.5\sim 900$ GeV 
\cite{ammo72,mors77,bari74,fire74,brom73,whit74,brea84,alne84,alne87,anso89}. 
The method of an extrapolation has been utilized for the predictions.

On the other hand, recently Alexopoulos et al., the E735 Experiment at the 
FNAL \cite{alex98} has reported the multiplicity distributions of the full 
phase space at 300, 546, 1000 and 1800 GeV. The data corrected by the 
simulation program are shown as the data of full phase space. Thus it is 
possible to compare our predictions with the multiplicity distributions at the 
Tevatron collider. Indeed we are very interested in these comparisons, because 
we can confirm whether or not the hadronization process is ruled by the 
two-component stochastic model.

In the next paragraph, the two-component stochastic model is introduced. In 
the 3rd paragraph, we compare the predictions with observed data by E735 
at the Tevatron at $\sqrt s = 1.8$ TeV. In the 4th one, we consider why there 
are discrepancies among the predictions and the data by E735 at the Tevatron. 
In the final one, we present our concluding remarks.

%
\section{Two-component stochastic model}
We explain briefly the essentials of the two-component stochastic model used 
in ref. \cite{mizo93}. It is  based on the following two-component branching 
equation for the two-component 
probability $P(n_a, n_b; t)$:
\begin{eqnarray}
  \frac{\partial P(n_a, n_b; t)}{\partial t} &=& \mu \left[ P(n_a-1, n_b; t) - 
  P(n_a, n_b; t) \right] \nonumber\\
  &&+ \lambda \left[(n_b-1)P(n_a, n_b-1; t) - n_b P(n_a, n_b; t) \right].
  \label{eqn:a1}
\end{eqnarray}
In eq. (\ref{eqn:a1}) the Poisson component (type $a$ particles) and the 
pure-birth (PB) component (type $b$ particles) do not couple to each other: 
The former is mainly related to particles  in the fragmentation region whereas 
the latter is related to those in the central region. The two-component 
probability $P(n_a, n_b; t)$ is therefore a product of multiplicity 
distributions for the Poisson component, $P_a(n_a; t)$ and the corresponding 
one for the PB component, $P_b(n_b; t)$: 
\begin{equation}
  P(n_a, n_b; t) = P_a(n_a; t)P_b(n_b; t),
  \label{eqn:a2}
\end{equation}
and the multiplicity distribution in the final state (i.e., for maximum $t = 
T$ (the maximum value of an evolution time)) is
\begin{equation}
  P(n) = \sum_{n_a+n_b=n} P(n_a, n_b; T),
  \label{eqn:a3}
\end{equation}
%
%
\begin{figure}[htb]
  \centering
  \epsfig{file=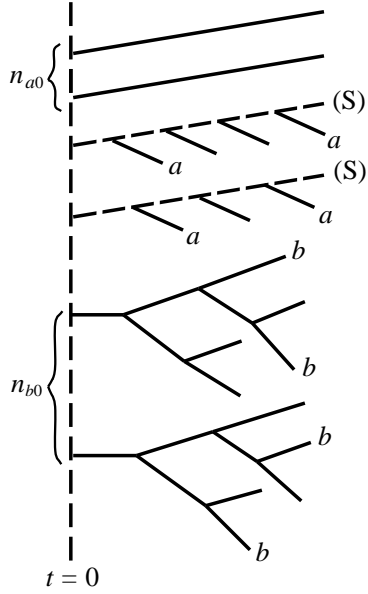,scale=0.45}
  \caption{Illustration of the two-component stochastic model: The dashed line 
  denotes the virtual source $(S)$ which cannot be observed. The particles 
  $a$ are produced by the Poisson process, and $b$ by the PB process. }
  \label{fig1}
\end{figure}
Here we use the following initial condition \cite{biya91}:
\begin{equation}
  P(n_a, n_b; t=0) = \frac{n_{a0}^{n_a}}{n_a !} e^{-n_{a0}} 
  \frac{n_{b0}^{n_b}}{n_b !} e^{-n_{b0}},
  \label{eqn:a4}
\end{equation}
which leads to 
\begin{equation}
  P_a(n_a, \langle n_a \rangle) = \frac{\langle n_a \rangle^{n_a}}{n_a!} 
  e^{- \langle n_a \rangle}, 
  \label{eqn:a5}
\end{equation}
and 
\begin{equation}
  P_b(n_b, \langle n_b \rangle) = \left\{
  \begin{array}{ll}
    e^{-n_{b0}} & \mbox{     for   } n_b = 0 \\
    \noalign{\vskip0.1cm}
    \displaystyle{\frac{\langle n_b \rangle}{n_b} 
    \frac{p^{n_b-1}}{(1+p)^{n_b+1}}} e^{- n_{b0}} L_{n_b-1}^{(1)} \left( - 
    \displaystyle{\frac{n_{b0}}p} \right) & \mbox{     for    } n_b \geq 1,
  \end{array} \right.
  \label{eqn:a6}
\end{equation}
where $p=e^{\lambda T}-1$. Here and after $L^{(k)}_{n}(-x)$ denotes the 
normalized Laguerre polynomials. The total multiplicity distribution is 
therefore given by
\begin{eqnarray}
  P(n) &=& \sum_{n_a+n_b=n} P_a(n_a, \langle n_a \rangle) P_b(n_b, \langle n_b 
  \rangle) \nonumber \\
  &=& e^{-(\langle n_a \rangle + n_{b0})} 
  \left[ \frac{\langle n_a \rangle^n}{n!} + \sum_{j=1}^n \frac{\langle n_a 
  \rangle^{n-j}}{(n-j)!} \left( \frac p{1+p} \right)^j \frac 1j \frac{n_{b0}}p 
  L_{j-1}^{(1)} \left( - \displaystyle{\frac{n_{b0}}p} \right) \right].
  \label{eqn:a7}
\end{eqnarray}
The corresponding factorial moments for this $P(n)$ are given by
\begin{eqnarray}
  F^{(l)} &=& \langle n(n-1) \cdots (n-l+1) \rangle \nonumber \\
  &=& \langle n_a \rangle^l + \sum_{j=1}^l {l\choose j} \langle n_a 
  \rangle^{l-j} \Gamma (j) \langle n_b \rangle p^{j-1} L_{j-1}^{(1)} 
  \left(- \frac{\langle n_b \rangle}p \right).
  \label{eqn:a8}
\end{eqnarray}
The $C_q$ moments which we calculate in what follows are easily derived  from 
$F^{(l)}$. We use the same parameters: $\mu T$, $n_{a0}$, $\lambda T$ and 
$n_{b0}$ \cite{mizo93} which were determined by fits to data for energies 
$\sqrt s = 11.5\sim 900$ GeV 
\cite{ammo72,mors77,bari74,fire74,brom73,whit74,brea84,alne84,alne87,anso89} 
by choosing minimum chi-squared of sum of $C_q\: (q=2\sim 5)$. Finally we 
have the following expressions:
\begin{eqnarray}
  \frac{\langle n_b \rangle}{n_{b0}} &=& \exp(\lambda T) = \exp \left(\lambda 
  \ln \sqrt{s/s_0} \right),
  \label{eqn:a9} \\
  \langle n_a \rangle &=& \mu T +n_{a0},
  \label{eqn:a10}
\end{eqnarray}
with $n_{b0} = 4.2$, $\lambda = 0.42 \pm 0.01$, $\sqrt{s_0} = 8.60 \pm 0.33$ 
GeV, $\mu = 1.22 \pm 0.10$ and $n_{a0} = 1.50 \pm 0.12$.

%
\section{Comparisons of predictions with data at $\sqrt s = 1.8$ TeV}
\begin{figure}[htb]
  \centering
  \epsfig{file=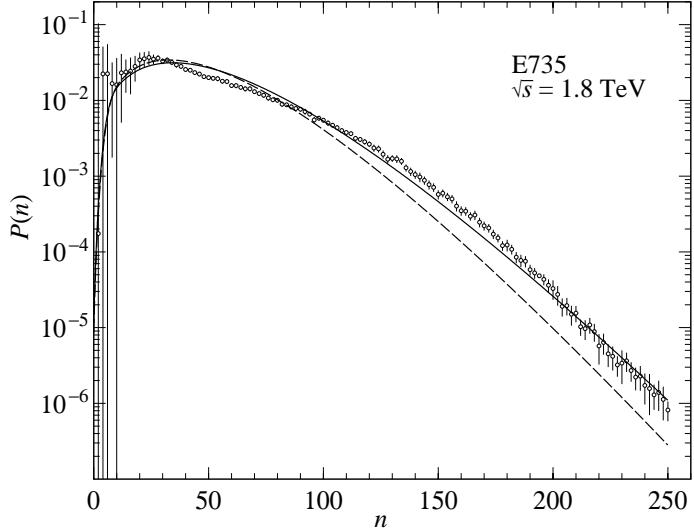,scale=0.45}
  \caption{Prediction for multiplicity by eqs. (\ref{eqn:a9}) and 
  (\ref{eqn:a10}) and data at the Tevatron. The solid line is obtained by 
  $\langle n\rangle = 47.88 + 1.54$ and $C_2 = 1.316 + 0.007$. The dashed line 
  is obtained by $\langle n\rangle = 47.88 - 1.54$ and $C_2 = 1.316 - 0.007$.}
  \label{fig2}
\end{figure}

Using eqs. (\ref{eqn:a9}) and (\ref{eqn:a10}), we can predict $\langle n 
\rangle$ and $C_q = \langle n^2 \rangle/\langle n \rangle^2$ in Table 
\ref{table1} and Fig. \ref{fig2}. Combining the data at $\sqrt s = 1.8$ TeV in 
ref. \cite{alex98} and our predictions, we find that there are discrepancies 
among our predictions and the data.
%
%
\begin{table}[htb]
  \centering
  \caption{Prediction for multiplicity by eqs. (\ref{eqn:a9}) and 
  (\ref{eqn:a10}) and corrected data of the full phase space at the Tevatron}
  \label{table1}
  \begin{tabular}{ccc}
  \hline
   & $\langle n\rangle$ & $C_2 = \langle n^2\rangle/\langle n\rangle^2$\\
  \hline
  prediction \cite{mizo93} & $47.88\pm 1.54$ & $1.316\pm 0.007$\\
  corrected data at $\sqrt s = 1.8$ \cite{alex98} &  $45.81\pm 0.77$ 
  & $1.45\pm 0.05$\\
  \hline
  \end{tabular}
\end{table}
%
%

\begin{figure}[htb]
  \centering
  \epsfig{file=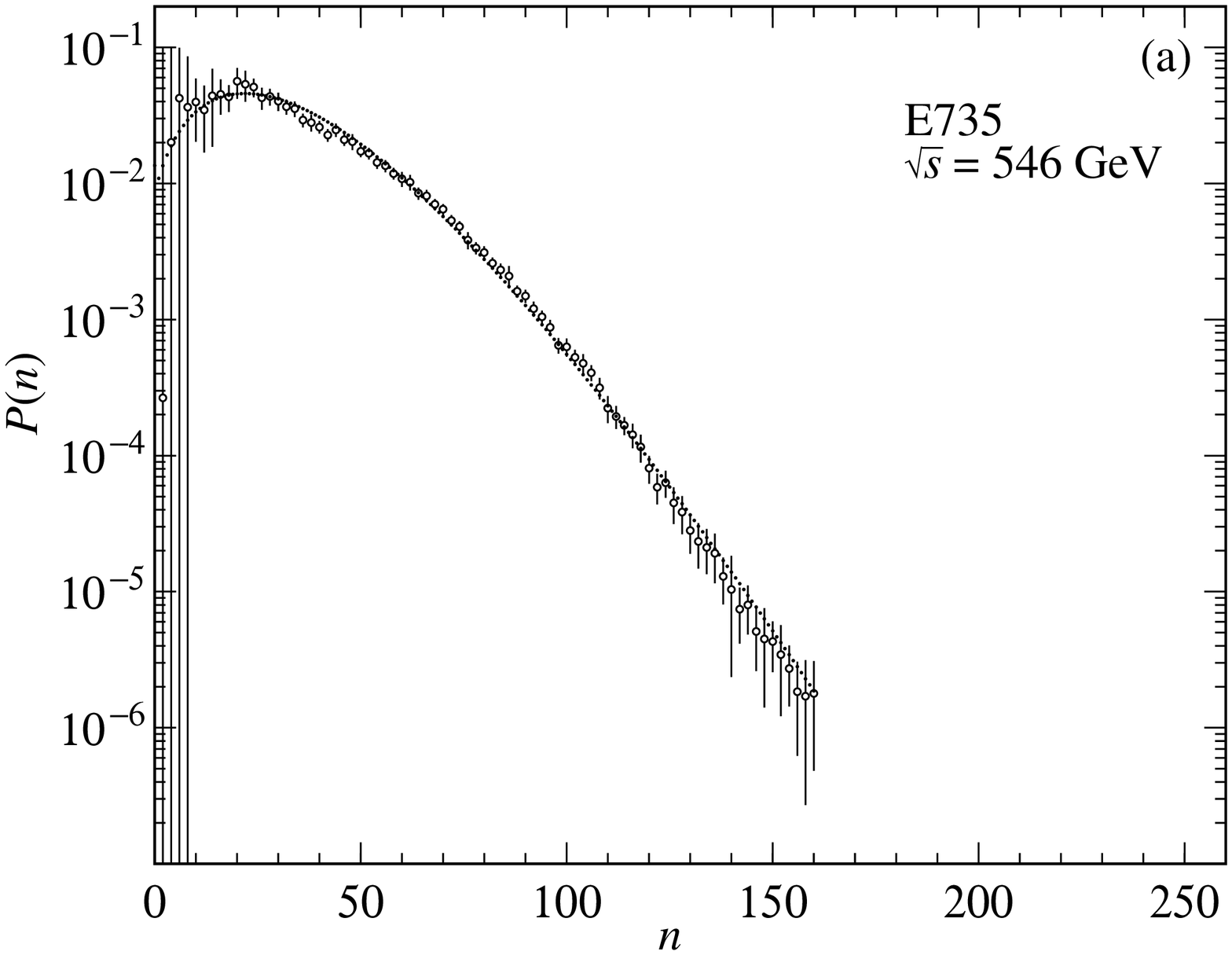,scale=0.45}
  \epsfig{file=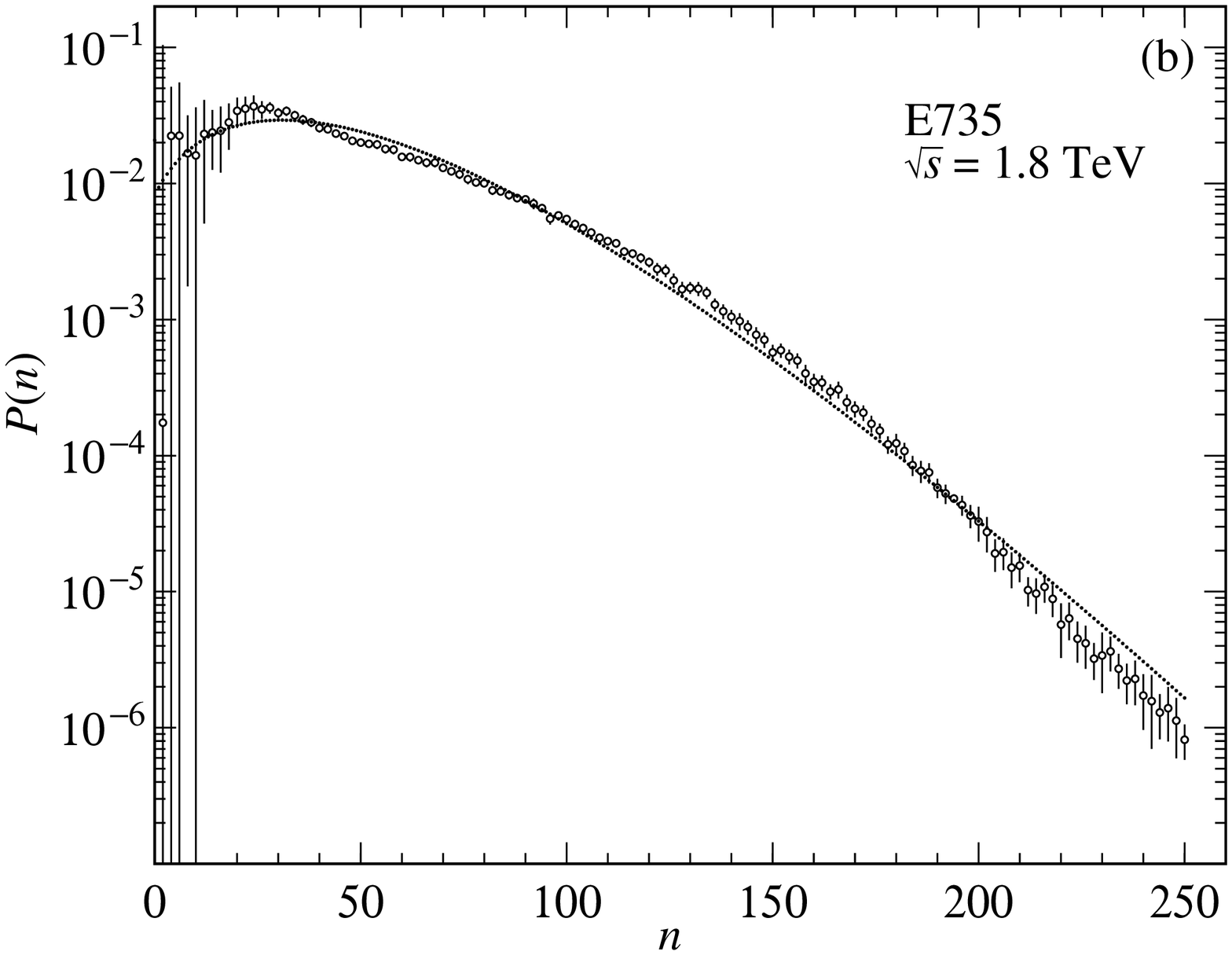,scale=0.45}
  \caption{(a) Analyses of data at $\sqrt s = 546$ GeV by eq. (\ref{eqn:a7}) 
  with the CERN MINUIT program. $\chi^2/N_{dof} = 49.2/77$. (b) Analyses of 
  data at $\sqrt s = 1.8$ TeV. $\chi^2/N_{dof} = 330.6/122$.}
  \label{fig3}
\end{figure}

To elucidate the reasons, we adopt other ways. Using eq. (\ref{eqn:a7}) with 
the CERN-MINUIT program, and the multiplicity distributions at the FNAL, the 
CERN S\= ppS and the Tevatron, we obtain various results given in Table 
\ref{table2}. See Fig. \ref{fig3}.
%
%
\begin{table}[htb]
  \centering
  \caption{Analyses of data \cite{ammo72,mors77,bari74,fire74,brom73,whit74,
brea84,alne84,alne87,anso89,alex98} by eqs. (\ref{eqn:a7}) $\sim$ 
(\ref{eqn:a10}) with the CERN MINUIT program. $\langle n \rangle = \langle n_a 
\rangle + \langle n_b \rangle$ with $\langle 
n_b \rangle = \frac12 n_{b0} [1+\sqrt{1+2 \langle n \rangle^2 (C_2 -1 -1/ 
\langle n \rangle )/n_{b0}} ]$. The data by UA5 \cite{alne84,alne87,anso89} 
are arranged by means of $\sigma_{\tiny\rm NSD} = \sigma_{\tiny\rm ND} + 
\sigma_{\tiny\rm DD}$, where NSD, ND and DD stand for ``non-single 
diffractive'', ``non-diffractive'' and ``double diffractive'', respectively.}
  \label{table2}
  \begin{tabular}{cccccc}
  \hline
  Exps. &  $n_{b0}$ & $\langle n\rangle$ & $C_2$ & $\langle n_a\rangle$ 
  & $\chi^2/N_{dof}$\\
  \hline
  FNAL 300 GeV &  5.301$\pm$0.423 & 8.433 & 1.259$\pm$0.028 & 0 & 24.2/11\\
  FNAL 800 GeV &  5.370$\pm$0.053 & 10.16 & 1.274$\pm$0.004 & 0 & 48.1/13\\
  ISR 30.4 GeV &  2.711$\pm$0.611 & 10.58$\pm$0.13 & 1.186$\pm$0.006 & 5.259 
  & 19.5/14\\
  ISR 44.5 GeV &  2.744$\pm$0.541 & 12.11$\pm$0.11 & 1.198$\pm$0.007 & 5.731 
  & 5.12/16\\
  ISR 52.6 GeV &  4.366$\pm$0.677 & 12.73$\pm$0.10 & 1.206$\pm$0.005 & 3.487 
  & 4.73/18\\
  ISR 62.2 GeV &  6.540$\pm$1.417 & 13.67$\pm$0.129 & 1.194$\pm$0.005 & 1.212 
  & 22.5/17\\
  UA5 200 GeV  &  3.419$\pm$0.592 & 21.46$\pm$0.29 & 1.249$\pm$0.010 & 7.005 
  & 9.58/28\\
  UA5 546 GeV  &  3.360$\pm$0.180 & 29.52$\pm$0.18 & 1.277$\pm$0.004 & 8.893 
  & 53.0/44\\
  UA5 900 GeV  &  2.703$\pm$0.160 & 35.93$\pm$0.33 & 1.292$\pm$0.007 & 13.08 
  & 55.6/51\\
  E735 300 GeV &  5.963$\pm$0.018 & 25.97 & 1.297$\pm$0.001 & 0 & 65.8/57\\
  E735 546 GeV &  4.993$\pm$0.020 & 31.06 & 1.368$\pm$0.002 & 0 & 49.2/77\\
 E735 1000 GeV &  4.699$\pm$0.020 & 38.96 & 1.400$\pm$0.002 & 0 &112.9/74\\
 E735 1800 GeV &  4.565$\pm$0.002 & 46.36 & 1.417$\pm$0.000 & 0 &330.6/122\\
  \hline
  \end{tabular}
\end{table}
%
%

Table \ref{table2} shows that the data at the Tevatron do not contain the 
multiplicity by the Poisson process in eq. (\ref{eqn:a7}), because $\langle 
n_a\rangle = 0$. On the other hand, the leading particle effect (finite 
$\langle n_a\rangle$) can be observed in the data by UA5 and in the ISR 
regions.

%
\section{Differences between data by UA5 and those at Tevatron}
To consider the reason of $\langle n_a \rangle = 0$ at the Tevatron, we 
analyse the data at $\sqrt s = 546$ GeV and $\sqrt s = 1.8$ TeV by CDF 
Collaboration \cite{abe94}. What we expect is small $\langle n_a \rangle$'s, 
because of no correction by the simulation program in ref. \cite{abe94}. 
Indeed we find that the magnitude of $\langle n_a\rangle$ is small, due to the 
restriction of the pseudo-rapidity cutoff $|\eta|<3.25$. This fact suggests 
that the data by CDF mainly produced in the central region. (See also 
\cite{ruso01}.)
%
%
\begin{table}[htb]
  \centering
  \caption{Analyses of data \cite{abe94} by eqs. (\ref{eqn:a7}) $\sim$ 
  (\ref{eqn:a10}). Data are read by eye-balls from their Fig. 10.}
  \label{table3}
  \begin{tabular}{cccccc}
  \hline
  Exps. &  $n_{b0}$ & $\langle n\rangle$ & $C_2$ & $\langle n_a\rangle$ 
  & $\chi^2/N_{dof}$\\
  \hline
  CDF 546 GeV  &  6.708$\pm$0.348 & 36.03$\pm$0.21 & 1.242$\pm$0.004 & 1.967 
  & 201.2/43\\
  CDF 1.8 TeV  &  4.754$\pm$0.150 & 45.22$\pm$0.40 & 1.358$\pm$0.007 & 2.344 
  & 157.0/64\\
  \hline
  \end{tabular}
\end{table}

As the next step, we carefully compare the data at $\sqrt s = 546$ GeV by UA5 
and E735. We find that the comparison of the data ($\sqrt s = 546$ GeV) 
of UA5 and E735 shows us that error bars of low multiplicities 
$\delta P(n)\ (n=2,\, 4,\, 6)$ at the Tevatron are larger than those by UA5. 
Then we can examine the above observation by making a pseudo-multiplicity at 
$\sqrt s = 546$ GeV as follows: The magnitude of error bars $\delta P(n)$ at 
$n=2,\ 4$ are assumed to be $\delta P(n)\sim P(n)\ (n=2,\ 4)$ in Table 
\ref{table4}, to consider our problem. As the same situation is seen in 
$P(n)$ at $\sqrt s = 1.8$ TeV in Table \ref{table5}
\footnote{
We have confirmed that the multiplicity distribution derived from the 
generalized gamma distribution in QCD \cite{doks93} can explain data with 
smaller chi-squared values than those of Table \ref{table2}. However, it is 
difficult to find physical meanings of a set of parameters:
$$
P_k(n,\: \langle n\rangle,\: D) = \frac{\langle n\rangle^n}{n!}N
\int_0^{\infty} z^{n+\mu k -1}e^{-z\langle n\rangle - D^{\mu}z^{\mu}} dz\: ,
$$
where $z = n/\langle n\rangle$, $N = \mu D^{\mu k}/\Gamma (k)$ and $k = 1/2$ 
($\chi^2/N_{dof} = 88.8/123$ at $\sqrt s = 1.8$ TeV). (See also refs. 
\cite{hegy00,biya84}.)
}, 
we make two kinds of pseudo-multiplicity distributions and analyse them.
%
%
\begin{table}[htb]
  \centering
  \caption{$P(n)$ and pseudo-$P(n)$ with $\delta P(n)\sim P(n)$ $(n = 2,\ 4)$ 
at $\sqrt s = 546$ GeV}
  \label{table4}
  \begin{tabular}{cccc}
  \hline
   & $P(n)$ & $P(n)$ & pseudo-$P(n)$ with \\
   & $\sqrt s = 546$ GeV & $\sqrt s = 546$ GeV & $\delta P(n)\sim 
P(n)\ (n=2,\ 4)$\\
   & UA5 & E735 Experiment & $\sqrt s = 546$ GeV E735\\
 $n$ & (real $P(n)\pm \delta P(n)$) & (real $P(n)\pm \delta P(n)$) 
& (pseudo-$P(n)\pm \delta P(n)$)\\
  \hline
  $2$ & $0.0027\pm 0.0008$ & $2.66\times 10^{-4}\pm 0.0997$ 
  & $2.66\times 10^{-4}\pm 2.66\times 10^{-4}$\\
  $4$ & $0.0077\pm 0.0013$ & $0.0201\pm 0.0799$ & $0.0201\pm 0.0201$\\
  $6$ & $0.0122\pm 0.0014$ & $0.0423\pm 0.0571$ & the same as left column\\
  $8$ & $0.0195\pm 0.0017$ & $0.0364\pm 0.0496$ & $\vdots$\\
  $\vdots$ & $\vdots$ & $\vdots$\\
  \hline
  $\chi^2/N_{dof}$ & $53.0/44$ & $49.2/77$ & $80.3$\\
  $(\langle n_a\rangle,\: \langle n_b\rangle)$ & $(8.89,\: 20.63)$ 
  & $(0,\: 31.06)$ & $(8.09,\: 24.17)$\\
  $\langle n\rangle$ & $29.52\pm 0.18$ & $31.06$ & $32.26\pm 0.32$\\
  $C_2$ & $1.277\pm 0.004$ & $1.368\pm 0.002$ & $1.304\pm 0.007$\\
  \hline
  \end{tabular}
\end{table}
%
%
\begin{table}[htb]
  \centering
  \caption{The same as Table \ref{table4} but $\sqrt s = 1.8$ TeV}
  \label{table5}
  \begin{tabular}{cccc}
  \hline
   & P(n) & 1st pseudo-$P(n)$ & 2nd pseudo-$P(n)$ with \\
   & $\sqrt s = 1.8$ TeV & $P(2) = P(4) = 0$ 
   & $\delta P(n)\sim P(n)\ (n=2,\ 4)$\\
   & E735 Experiment & $\sqrt s = 1.8$ TeV E735 
   & $\sqrt s = 1.8$ TeV E735\\
  $n$ & (real $P(n)\pm \delta P(n)$) & (pseudo-$P(n)\pm \delta P(n)$) 
  & (pseudo-$P(n)\pm \delta P(n)$)\\
  \hline
  $2$ & $1.75\times 10^{-4}\pm 0.1038$ & $0$ 
  & $1.75\times 10^{-4}\pm 1.75\times 10^{-4}$\\
  $4$ & $0.0223\pm 0.0291$ & $0$ & $0.0223\pm 0.0223$\\
  $6$ & $0.0224\pm 0.0327$ & the same as left column 
  & the same as left column\\
  $8$ & $0.0167\pm 0.0150$ & $\vdots$ & $\vdots$\\
  $\vdots$ & $\vdots$\\
  \hline
  $\chi^2/N_{dof}$ & $330.6/122$ & $314.8/120$ & $432.3/122$\\
  $(\langle n_a\rangle,\: \langle n_b\rangle)$ & $(0,\: 46.36)$ 
  & $(0,\: 46.82)$ & $(7.94,\: 39.26)$\\
  $\langle n\rangle$ & $46.36$ & $46.82$ & $47.21\pm 0.19$\\
  $C_2$ & $1.417\pm 0.000$ & $1.410\pm 0.000$ & $1.372\pm 0.003$\\
  \hline
  \end{tabular}
\end{table}

The results of Tables \ref{table4} and \ref{table5} show that $\langle n_a 
\rangle$ and $C_2$ depend on the magnitudes of error bars at $n =$ 2 and 4, as 
is shown by analysing the pseudo-$P(n)$'s. The low multiplicities with larger 
error bars are equal to $P(2) = P(4) = 0$, as is seen in the first 
pseudo-$P(n)$ in Table \ref{table5}. We find that $\langle n_a \rangle$ from 
the second pseudo-$P(n)$ in Table \ref{table5} is coincided with the 
prediction of Table \ref{table1}. Thus, it can be said that the leading 
particle effect is relating to low multiplicities and the magnitude of error 
bars.

%
\section{Concluding remarks}
We have compared our predictions \cite{mizo93} for the multiplicity 
distribution at 1.8 TeV, based on the two-component model, and the data at the 
Tevatron by E735. It is found that there are discrepancies between 
them. See Fig. \ref{fig2} and Table \ref{table1}.

To elucidate physical reason of the discrepancies, we have analysed the data 
by means of the two-component model. From this procedure, we have known that 
the data by UA5 contain the reading particle effect, finite $\langle n_a 
\rangle$. On the other hand, the data by the Tevatron do not contain it
\footnote{
It should be noticed that the streamer chamber at the CERN S\= ppS and the 
tracking detector at the Tevatron are used. The pseudo-rapidity range of the 
Tevatron is $|\eta| < 3.25$, which corresponds to the central region. As 
events of low multiplicity probably contain particles with larger $\eta$, it 
may be said that the leading particle might be missed in measurements at the 
Tevatron. Of course, the corrected data of the full phase space should contain 
the effect.
}. 

As the next step, we have resolved the reason why the data by the Tevatron do 
not contain $\langle n_a \rangle$. For this purpose, we analyse the data by 
CDF \cite{abe94} and find small $\langle n_a\rangle$ at $\sqrt s = 546$ GeV 
and 1.8 TeV. See Table \ref{table3}. Moreover, we compare the data by UA5 and 
at the Tevatron in Tables \ref{table4} and \ref{table5} and find that the 
error bars of low multiplicities, $\delta P(n)$ ($n=2,\ 4,\ 6$) are large. To 
see how our result depends on large $\delta P(n)$ ($n=2,\ 4,\ 6$), we have 
made the pseudo-multiplicity distribution, assuming $\delta P(n) \approx P(n)$ 
at $n=2$ and $4$. From the pseudo-multiplicity distribution, we can estimate 
the finite leading particle effect at $\sqrt s = 546$ GeV and 1.8 TeV at the 
Tevatron. The result of $\langle n_a\rangle \approx 8$ from the second 
pseudo-$P(n)$ in Table \ref{table5} is almost the same of our prediction 
\cite{mizo93}. (See Table \ref{table1}.)

In conclusion, we have found that the corrected data of full phase space at 
Tevatron by E735 do not contain the leading particle effect which 
is observed in the data by UA5. It is difficult to estimate this effect, 
because of larger error bars at low multiplicities, $P(n)\pm \delta P(n)$ 
($n=2,\ 4$)
\footnote{
For us, it is difficult to discuss the valuation of the correction methods 
applied to the data at the Tevatron by E735.
}. 
In other words, the low multiplicities with $\delta P(n) > P(n)$ $(n=2,\ 4)$ 
are equivalent to $P(n) = 0$ $(n=2,\ 4)$. They are playing important roles for 
determining the leading particle effect which is expected in ref. 
\cite{mizo93}.

\begin{figure}[htb]
  \centering
  \epsfig{file=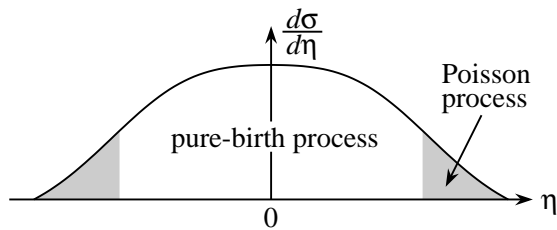,scale=0.6}
  \caption{$d\sigma/d\eta$ distribution. The PB and the Poisson processes 
  probably rule the central and fragmentation regions, respectively.}
  \label{fig4}
\end{figure}

Moreover, providing $P(n,\ \delta \eta)$ with the pseudo-rapidity cutoff in a 
future, we could obtain more useful information for the production mechanism 
which are illustrated in Fig. \ref{fig4}: The pure-birth (PB) process mainly 
describes $P(n)$ in the central region. On the other hand, the Poisson process 
mainly does $P(n)$ in the fragmentation region
\footnote{
Two papers relating to $d\sigma/d\eta$ are presented in different points of 
view by Takagi and Tsukamoto \cite{taka98}, and Ohsawa \cite{ohsa94}.
}. 
%
%

%
\section*{Acknowledgements}
We are greatly appreciated Prof. W. D. Walker for his sending the data at the 
Tevatron collider by E735 Experiment. One of authors (M.B.) is indebted to 
Japanese Grant-in-aid for Education, Science, Sports and Culture 
(No. 09440103).

%

%
\end{document}